\documentclass[a4paper,12pt]{article}

\usepackage{amsmath,amsfonts,amsthm,amssymb,dsfont}
\usepackage{graphicx,wrapfig,lipsum}
\usepackage{float}
\usepackage{multicol}
\usepackage{braket}
\usepackage{color}
\usepackage{cite}
\usepackage{MnSymbol,wasysym}
\usepackage{hyperref}
\usepackage{subfigure}
\usepackage{tikz}
\usetikzlibrary{shapes.geometric}
\newcommand{\floor}[1]{\left\lfloor #1 \right\rfloor}
\numberwithin{equation}{section}

\begin{document}


\begin{center}
\noindent{\textbf{\LARGE{Comments on non-Abelian T-duals and their holographic description}}}
\smallskip
\smallskip

\smallskip
\smallskip

\smallskip
\smallskip

\smallskip
\smallskip

\noindent{Paul Merrikin\footnote{paulmerrikin@hotmail.co.uk}, Ricardo Stuardo\footnote{ricardostuardotroncoso@gmail.com}}

\smallskip
\smallskip

\smallskip
\smallskip

\smallskip
\smallskip

\smallskip
\smallskip

\smallskip
\smallskip

\smallskip

{Department of Physics, Swansea University, Swansea SA2 8PP, United Kingdom}


\end{center}
\smallskip
\smallskip
\smallskip
\smallskip
\abstract
We present three different Type IIB AdS$_4$ Supergravity solutions, derived using the electrostatic-like problem formalism which preserves 8 Poincar\'{e} Supercharges.
Two of these solutions correspond to the Abelian and non-Abelian T-duals of a Type IIA background, obtained from the dimensional reduction of 11D AdS$_4 \times S^7$ Supegravity. The third solution is a new background with some special properties. We compute the Page charges in each case, and provide brane set-up descriptions of the geometries. Also, we argue that the non-Abelian T-dual can be understood as a zoom-in of a more general background with a well defined holographic dual.

\smallskip
\smallskip

\smallskip



\newpage
\newpage\tableofcontents
\newpage

\section{Introduction}

The AdS/CFT conjecture \cite{Maldacena:1997re} allows us to give geometrical descriptions of super-conformal field theories in different dimensions in terms of a given brane configuration in Supergravity. For the case of $\mathcal{N}=4$ Supersymmetric theories, these gauge/gravity dualities had been established in different dimensions, e.g. \cite{Lozano:2020txg}-\cite{Lozano:2021rmk} for AdS$_{2}$, \cite{Macpherson:2018mif}-\cite{Lozano:2019jza} for AdS$_{3}$,\cite{DHoker:2007zhm}-\cite{Akhond:2021ffz} for AdS$_{4}$, \cite{Gaiotto:2009gz}-\cite{Aharony:2012tz} for AdS$_{5}$, \cite{DHoker:2016ujz}-\cite{Legramandi:2021aqv} for AdS$_{6}$ and \cite{Apruzzi:2013yva}-\cite{Nunez:2018ags} for AdS$_{7}$ among others.

When considering the Abelian T-duality (ATD) transformation of a Supergravity solution with a certain holographically dual field theory, since it is a symmetry of string theory, both geometries describe the same dynamics in different ways. Therefore both geometries are holographically dual to the same field theory.

The previous statement is no longer valid if one considers non-Abelian T-duality (NATD) transformations. These transformations, firstly developed for pure NS-NS fields in \cite{delaOssa:1992vci} and then extended for backgrounds containing R-R fields that have a $SU(2)$ isometry subgroup in \cite{Sfetsos:2010uq} (the extension beyond $SU(2)$ was done in \cite{Lozano:2011kb} and an explicit flux transformation was given in \cite{Kelekci:2014ima}), do not preserve the symmetries of the compact manifold along which one performs the transformations. Hence, the background obtained after the NATD transformations describes different dynamics with respect to its seed, which can be seen at the level of the string $\sigma$-model \cite{Giveon:1993ai}. In this manner NATD is not a symmetry of String Theory but rather a solution generating method, see e.g. \cite{Gasperini:1993nz,Elitzur:1994ri}. 

It is then natural to ask what is the corresponding dual field theory of a background obtained after NATD transformations. This was first addressed in \cite{Itsios:2012zv,Itsios:2013wd}, where it was shown that it is indeed possible to describe properties of the NATD geometry in terms of those of its predecessor, e.g. observables charged under the dualised isometries changed after the transformation while non-charged (neutral) observables and the holographic central charge remain the same. However, no precise proposal for a dual field theory to NATD geometries was given until the work of \cite{Lozano:2016kum} in AdS$_5 \times S^{5}$, where the dual theory came in the form of a `long quiver' CFT preserving $\mathcal{N}=2$ supersymmetry.

In \cite{Lozano:2016wrs} the NATD of the AdS$_{4}\times S^{7}$ reduction to Type IIA Supergravity was shown to have a brane-configuration given by an unbounded Hanany-Witten set-up \cite{Hanany:1996ie}, which suggests that the dual field theory is not well defined. Using the formalism of \cite{Lozano:2016kum,Assel:2011}, a consistent way of completing the field theory was then developed, such that the completed form can be written in terms of a linear quiver. This leads to a new Type IIB background, in which the NATD appears as a zoom-in on a region of this completed background. 

Here we continue with the discussion about the holographic description of $\text{AdS}_4$ backgrounds obtained via NATD using the electrostatic-like problem set-up developed in \cite{Akhond:2021ffz}, providing an interpretation of NATDs as a zoom-in on backgrounds that have a well defined holographic dual, along similar lines to the $\text{AdS}_6$ case in \cite{Legramandi:2021uds}. 

The outline of the paper is as follows: in section 2 we present two solutions to Type IIB Supergravity on AdS$_{4}$, which correspond to the ATD and the NATD found in \cite{Lozano:2016wrs}, computing the respective Page charges in each case. In section 3 we argue how the NATD appears as a zoom-in of the backgrounds found in \cite{Akhond:2021ffz}, which were shown to be dual to linear quiver field theories. In section 4 we study a new AdS$_{4}$ Supergravity solution. We conclude in section 5 with a summary of the paper and some final remarks.

\section{AdS$_{4}$ Type IIB Background}

In \cite{DHoker:2007zhm}, an infinite family of solutions to Type IIB Supergravity that are dual to $\mathcal{N}=4$ super-conformal field theories were proposed. It was then showed in \cite{Akhond:2021ffz} that the same family of solutions can be obtained in the `electrostatic' formalism. In order to match the conformality and global symmetries of the field theory, the Supergravity background geometry must contain an AdS$_{4}$ factor and a couple of two spheres $S^{2}_{1}$ and $S^{2}_{2}$. The other NS and R background fields must preserve the isometries of the geometry. The Type IIB AdS$_4 \times S^2 \times S^2$ has the following background   
    \begin{equation}\label{eqn:back1}
    \begin{gathered}
        ds_{10,st}^2=f_1(\sigma,\eta)\bigg[ds^2(\text{AdS}_4)+f_2(\sigma,\eta)ds^2(S_1^2)+f_3(\sigma,\eta)ds^2(S_2^2)+f_4(\sigma,\eta)(d\sigma^2+d\eta^2)\bigg],
        \\e^{-2\Phi}=f_5(\sigma,\eta),~~~~~B_2=f_6(\sigma,\eta)\text{Vol}(S_1^2), ~~~C_2=f_7(\sigma,\eta)\text{Vol}(S_2^2), ~~\tilde{C}_4=f_8(\sigma,\eta)\text{Vol}(\text{AdS}_4),
        \\f_1=\frac{\pi}{2}\sqrt{\frac{\sigma^3\partial_{\eta\sigma}^2V}{\partial_{\sigma}(\sigma \partial_{\eta}V)}},~~~~ f_2=-\frac{\partial_{\eta}V\partial_{\sigma}(\sigma \partial_{\eta}V)}{\sigma\Lambda},~~~~f_3=\frac{\partial_{\sigma}(\sigma \partial_{\eta}V)}{\sigma \partial_{\eta \sigma}^2V}, ~~~~ f_4=-\frac{\partial_{\sigma}(\sigma \partial_{\eta}V)}{\sigma^2\partial_{\eta}V},
        \\f_5=-16\frac{\Lambda \partial_{\eta}V}{\partial_{\eta \sigma}^2V},~~~~f_6=\frac{\pi}{2}\Bigg(\eta-\frac{\sigma \partial_{\eta}V\partial_\eta^2V}{\Lambda}\Bigg),~~~~f_7=-2\pi\Bigg(\partial_\sigma(\sigma V)-\frac{\sigma \partial_\eta V \partial_\eta ^2V}{\partial_{\eta \sigma}^2V}\Bigg),
        \\f_8=-\pi^2\sigma^2\bigg(3\partial_\sigma V + \frac{\sigma \partial_\eta V \partial_\eta^2V}{\partial_\sigma(\sigma\partial_\eta V)}\bigg), ~~~~ \Lambda = \partial_\eta V \partial_{\eta \sigma}^2V + \sigma\big((\partial_{\eta\sigma}^2V)^2+(\partial_\eta^2V)^2\big),
    \end{gathered}
    \end{equation}
with 
    \begin{equation}\label{eqn:back2}
        F_1=0,~~~~H=dB_2,~~~~~F_3=dC_2,~~~~~F_5=d\tilde{C}_4+*d\tilde{C}_4.
    \end{equation}
Notice that all functions $f_{1},...,f_{8}$ can be determined in terms of a single function $V(\sigma,\eta)$, and partial derivatives.

In order to be a solution of Type IIB Supergravity, $V$ must satisfy
    \begin{equation}\label{eqn:2dpde}
        \partial_\sigma(\sigma^2\partial_\sigma V(\sigma,\eta))+\sigma^2\partial_\eta^2 V(\sigma,\eta)=0.
    \end{equation}
By using the redefinition
    \begin{equation}
        V(\sigma,\eta)=\frac{\partial_\eta W(\sigma,\eta)}{\sigma},
    \end{equation}
equation \eqref{eqn:2dpde} becomes the 2D Laplace equation
    \begin{equation}\label{eqn:laplace}
        \partial_\sigma^2 W(\sigma,\eta)+\partial_\eta^{2}W(\sigma,\eta)=0.
    \end{equation}
Within this set-up, the boundary conditions imposed in \cite{Akhond:2021ffz} were analogous to an electrostatic problem in which there are two parallel conducting planes and a line of charge density given by $\mathcal{R}(\eta)$, which is usually called the `Rank function'. This Rank function, which must be linear by sections in order to have quantized Page charges, completely characterises the brane configuration and therefore the field content of the dual quiver theory.

We are interested in solutions of the form \eqref{eqn:back1} which do not necessarily satisfy the same boundary conditions. In the variables $z,\bar{z}=\sigma\pm i\eta$, any solution to the Laplace equation can be written as $W=f(z)+\bar{f}(\bar{z})$. In what follows we will only consider the case in which $f(z)$ and $\bar{f}(\bar{z})$ are regular, hence we can expand them in a Taylor series. After imposing a reality condition for $W$, namely $W=W^{\star}$, and going back to the variables $\sigma,\eta$, the most general solution to equation \eqref{eqn:laplace} (ignoring boundary conditions) takes the form
    \begin{equation}\label{eqn:W}
        W = \alpha_{0} + \sum^{+\infty}_{n=1}\left[ \alpha_{n}\sum^{\floor{\frac{n}{2}}}_{k=0} \binom{n}{2k} (-1)^{k} \sigma^{n-2k} \eta^{2k} + \beta_{n}\sum^{\floor{\frac{n-1}{2}}}_{k=0} \binom{n}{2k+1} (-1)^{k} \sigma^{n-2k-1} \eta^{2k+1} \right],
    \end{equation}
from which we obtain the following general solution to equation \eqref{eqn:2dpde}
    \begin{equation}\label{eqn:Veq}
        V = \sum^{+\infty}_{n=1}\left[ \alpha_{n}\sum^{\floor{\frac{n}{2}}}_{k=0} 2k\binom{n}{2k} (-1)^{k} \sigma^{n-2k-1} \eta^{2k-1} + \beta_{n}\sum^{\floor{\frac{n-1}{2}}}_{k=0} (2k+1)\binom{n}{2k+1} (-1)^{k} \sigma^{n-2k-2} \eta^{2k} \right].
    \end{equation}
We will now show that different solutions appear by considering truncations of the Taylor series.

\subsection{Abelian T-dual (ATD) Solution}

We first consider the case in which  all the coefficients $\alpha_{n}$ and $\beta_{n}$ are set to zero, except for $\alpha_{2}$ and $\alpha_{3}$. This then leads to the following potential
    \begin{equation}\label{eqn:A}
        V_\text{ATD}(\sigma,\eta) = 2\alpha_{2}\frac{\eta}{\sigma} -6\alpha_{3}\eta.
    \end{equation}
Let us now set $\alpha_{2}=6\alpha_{3}$, $\alpha_{3} = M/12$ and use the change of coordinates 
    \begin{equation}
        \sigma = 2\cos^{2}\left(\frac{\mu}{2}\right), \ \ \ \ \eta = \frac{2}{\pi}r.
    \end{equation}
The range of the coordinates are $\mu\in [0,\pi]$ and, as we will explain below, it is convenient to choose $r\in [n\pi,(n+1)\pi]$, where the points $r=n\pi$ and $r=(n+1)\pi$ are identified. With this parametrization, the background metric reads
    \begin{equation}\label{eqn:abelian}
        ds^{2} = \pi\cos\left(\frac{\mu}{2}\right)\left[ds^{2}(\text{AdS}_4) 
        + \sin^{2}\left(\frac{\mu}{2}\right)ds^{2}(S^{2}_{1})
        + \cos^{2}\left(\frac{\mu}{2}\right)ds^{2}(S^{2}_{2})
        + d\mu^{2}+\frac{4}{\pi^{2}\sin^{2}(\mu)}dr^{2}\right]    \end{equation}
and the rest of the background fields are given by
    \begin{align}
        e^{-2\Phi} &=4 M^{2}\tan^{2}\left(\frac{\mu}{2}\right),\\
        B_{2} &= r \, \text{Vol}(S^{2}_{1}), \\
        C_{2} &= 2M r\, \text{Vol}(S^{2}_{2}),\\
        \tilde{C}_{4} &= 6\pi M r\, \text{Vol}(\text{AdS}_{4}).
    \end{align}
The field strengths for the gauge forms are
    \begin{align}
        H_{3} &= dr \wedge\text{Vol}(S^{2}_{1}),\\
        F_{3} &=2M dr \wedge \text{Vol}(S^{2}_{2}) ,\\
        F_5 &= 6\pi M  dr\wedge\text{Vol}(\text{AdS}_4) - \frac{3}{4}\pi^{2}M\sin^3(\mu)d\mu \wedge \text{Vol}(S_{1}^{2}) \wedge \text{Vol}(S_{2}^{2}).
    \end{align}

By observation, one can easily see the presence of singularities at $\mu=\pi$ (from the D6 branes in Type IIA) and at $\mu=0$ (from the NS5 branes), as discussed in \cite{Lozano:2016wrs}, and hence the system is defined in the interval $\mu \in [0,\pi]$ (corresponding to $\sigma \in [0,1]$). The dilaton diverges to positive infinity as $\mu \rightarrow 0$ and to negative infinity as $\mu \rightarrow \pi$. Hence, the above Supergravity description is not well defined in these limits. 
   
This solution corresponds to the Abelian T-dual of the Type IIA solution obtained from a dimensional reduction of 11D Supergravity \cite{Lozano:2016wrs}. 

\subsubsection*{Page Charges}

We now calculate the Page Charges, using the following definition
    \begin{equation}\label{eqn:PageCharge}
        Q_{D_p/NS5}=\frac{1}{(2\pi)^{7-p}\alpha'}\int_{\Sigma_{8-p}}\widehat{F}_{8-p},
    \end{equation}
with $\alpha'=1$, $2\kappa_{10}^2T_{D_p}=(2\pi)^{7-p}$ and $\widehat{F}_p=F_p \wedge e^{-B_2}$. The background of equation \eqref{eqn:back1} has the following Page fluxes
    \begin{equation}\label{eqn:hats}
        \widehat{F}_3=F_3,~~~~~~~~~~~~~~~~~   \widehat{F}_5=F_5-B_2 \wedge F_3.
    \end{equation}

When integrating the charges, it is convenient to choose $r\in [n\pi,(n+1)\pi]$ since this is the minimal building block of the brane set-up, i.e. is the smallest size interval in which we get integer charges for all the gauge forms. \\\\
For the NS5 charge, we integrate in the submanifold $\mathcal{M}_{1}=(r,S^{2}_{1})$ for any value of $\mu$
    \begin{equation}
        N_{NS5}=\frac{1}{(2\pi)^2}\int_{\mathcal{M}_{1}}H_{3}=1.
    \end{equation}
For the D5 charge, the integration is performed in the three-cycle $\mathcal{M}_{2}=(r,S^{2}_{2})$ for any value of $\mu$
    \begin{equation}
        N_{D5}=\frac{1}{(2\pi)^2}\int_{\mathcal{M}_{2}}F_3=2M.
    \end{equation}
Finally, for the D3 charge, we integrate in the cycle $\mathcal{M}_{3}=(\mu,S^{2}_{1},S^{2}_{2})$ for any value of $r$
    \begin{align}
        N_{D3} &=-\frac{1}{(2\pi)^4}\int_{\mathcal{M}_{3}}\widehat{F}_5=M,
    \end{align}
where the negative charge suggests that the orientation of the cycle should be swapped in this case.  

If we had considered instead the interval $r\in [n\pi,(n+2)\pi]$, the NS5 and D5 charges would have counted the branes between $[n\pi,(n+1)\pi]$ and $[(n+1)\pi,(n+2)\pi]$, thus not giving the smallest possible building block.

\begin{figure}  
\centering  

\subfigure[Brane set-up]  
{  
\begin{tikzpicture}

\draw[->] (4,2) -- (4,2.5) node[above] {$\mu$};
\draw[->] (4,2) -- (4.5,2) node[right] {$r$};



\draw (2,1.5) node[above] {D5};

\node[label=above:] at (1.7,1.2){\LARGE $\otimes$};
\node[label=above:] at (2.3,1.2){\LARGE $\otimes$};


\draw (1,-2) -- (1,2.5);
\draw (1,2.6) node[above] {NS5};
\draw (1,-2.16) node[below] {$n\pi$};
\draw[dashed] (3,-2) -- (3,2.5);
\draw (3,2.6) node[above] {NS5};
\draw (3,-2) node[below] {$(n+1)\pi$};


\draw (1,0) -- (3,0);
\draw (2,-1) node[above] {D3};

\end{tikzpicture}
\label{fig:1BraneSetup}
}  
\quad\quad\quad\quad\quad\quad\quad
\subfigure[Linear quiver]  
{  

     \centering
 \begin{tikzpicture}[square/.style={regular polygon,regular polygon sides=4}]
        \node at (0,2) [square,inner sep=0.1em,draw] (f100) {$2M$};
        \node at (0,0) [circle,inner sep=0.5em,draw] (c100) {$M$};
        \draw (c100) -- (f100);
        \draw (0,-2) node[above] {};
        \draw (2,0) node[above] {};
        \draw (-2,0) node[above] {};
        \draw (0,-2.3) node[above] {};
        \draw  plot [smooth, tension=6] coordinates {(0.38,-0.38) (0,-1.5) (-0.38,-0.38)};
    \end{tikzpicture}
    \label{fig:Quiver11}
 
    \label{fig:Quiver1}
}

\caption{The Hanany-Witten (NS5, D3, D5) brane set-up for the ATD solution in (a), along with the corresponding balanced linear quiver diagram, (b). The brane set-up  consists of $M$ D3-branes stretched between NS5-branes located at $n\pi$ and $(n+1)\pi$, with $2M$ D5-branes in each interval (taking $M=1$). Since the points $r=n\pi$ and $r=(n+1)\pi$ are identified, the two NS5-branes in the figure correspond to the same brane (represented using a dashed line).}
\end{figure}
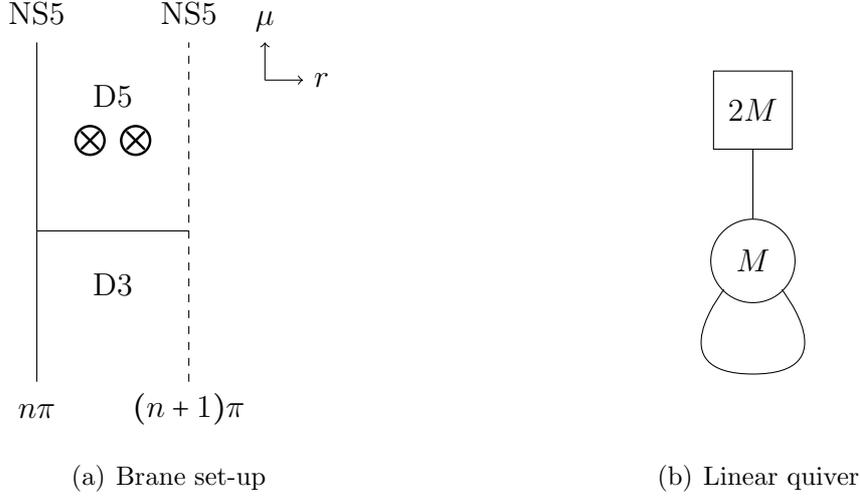

The full brane configuration for this geometry is described in figure
\ref{fig:1BraneSetup}, in which M D3-branes are stretched between NS5-branes located at $n\pi$ and $(n+1)\pi$, with 2M D5-branes in each interval. The two NS5-branes represent the same brane, as the points $n\pi$ and $(n+1)\pi$ are identified. The corresponding balanced linear quiver diagram is then given in figure \ref{fig:Quiver1}.


\newpage

\subsection{Non-Abelian T-dual (NATD) Solution}

Another well known solution can be obtained by considering the case in which only $\beta_{3}$ and $\beta_{4}$ are non-vanishing, namely
    \begin{equation}\label{eqn:NA}
        V_\text{NATD} = \beta_{3}\left( 3\frac{\eta^{2}}{\sigma} -3\sigma  \right) + \beta_{4}(4\sigma^{2}-12\eta^{2}).
    \end{equation}
    
After fixing $\beta_{3}= 8\beta_{4}$ and $\beta_{4}=M/96$, and using the change of coordinates
    \begin{equation}
        \sigma = 2\cos^{2}\left(\frac{\mu}{2}\right), \ \ \ \ \eta = \frac{2}{\pi}r,
    \end{equation}
we obtain the following metric
    \begin{equation}
        ds^{2} = \pi\cos\left(\frac{\mu}{2}\right)\left[ds^{2}(
        \text{AdS}_{4})
        +\frac{4}{\pi^{2}}\frac{r^{2}}{\Delta}\sin^{2}\left(\frac{\mu}{2}\right)ds^{2}(S^{2}_{1})
        +\cos^{2}\left(\frac{\mu}{2}\right)ds^{2}(S^{2}_{2})
        +d\mu^{2}+\frac{4}{\pi^{2}\sin^2(\mu)}dr^{2}\right],
    \end{equation}
where $\Delta = \Delta(\mu,r)$ is given by
\begin{equation}
    \Delta(\mu,r) = \sin^{2}\left(\frac{\mu}{2}\right)\sin^{2}(\mu)
            +\frac{4}{\pi^{2}}r^{2},
\end{equation}    
with the rest of the background fields given by
    \begin{align}
        e^{-2\Phi}&=M^2\tan^{2}\left(\frac{\mu}{2}\right) \Delta, \\
        B_{2} &=\frac{4}{\pi^{2}}\frac{r^{3}}{\Delta}\text{Vol}(S^{2}_{1}),\\
        C_{2} &= \frac{\pi}{4}M\left( \frac{4}{\pi^{2}}r^{2}-\cos(\mu)\left(\cos^{2}(\mu)-3\right) \right)\text{Vol}(S^{2}_{2}),\\
        \tilde{C}_{4} &= \frac{\pi^{2}}{8}M\left( \frac{24}{\pi^{2}}r^{2} -( \cos(2\mu)-4\cos(\mu) ) \right)\text{Vol}(\text{AdS}_{4}).
    \end{align}

The gauge fields lead to the following field strengths
    \begin{align}
        H_{3} &=  \frac{r^{2}}{2\pi^2\Delta^{2}}
        \left[ 8\left( \frac{4}{\pi^{2}}r^{2} + 3\sin^{2}\left(\frac{\mu}{2}\right)\sin^{2}(\mu) \right)dr 
        - r\bigg(\sin(\mu)+4\sin(2\mu)-3\sin(3\mu)\bigg) d\mu \right]
        \wedge \text{Vol}(S^{2}_{1}),\\
         F_{3} &= \frac{\pi}{2}M\left[ \frac{4}{\pi^{2}}r \, dr - \frac{3}{2}\sin^{3}(\mu) d\mu  \right]\wedge \text{Vol}(S^{2}_{2}),
    \end{align}
together with the self-dual 5-form
    \begin{equation}
    \begin{aligned}
        F_{5} &= M\frac{\pi^{2}}{8}
        d\left( \frac{24}{\pi^{2}}r^{2} -( \cos(2\mu)-4\cos(\mu) ) \right)\wedge\text{Vol}(\text{AdS}_{4}) 
        \\ &\phantom{=}-M
        \frac{r^{2}\sin^{2}(\mu)}{\pi\Delta}\left( 3r\sin(\mu) d\mu +2\sin^{2}\left(\frac{\mu}{2}\right)dr
        \right) \wedge\text{Vol}(S^{2}_{1})\wedge\text{Vol}(S^{2}_{2}).
    \end{aligned}
    \end{equation}

This solution reproduces the non-Abelian T-dual of the Type IIA solution, in \cite{Lozano:2016wrs}. As in the ATD, we have $\mu \in [0,\pi]$ due to the presence of singularities. Once again, the above description becomes weak when $\mu \rightarrow 0$ and $\mu \rightarrow \pi$, as the dilaton diverges in these limits.

\subsubsection*{Page Charges}

Now we compute the Page charges using equations \eqref{eqn:PageCharge}. Note that unlike the ATD, in this case, the region of $r$ is non-compact. However the charges below are calculated over the interval  $r \in [n\pi,(n+1)\pi]$, with $n \in \mathbb{N}$, since each one of these intervals is a minimal building block of the brane set-up.

Before computing the charges, it is convenient to perform the following large gauge transformation
    \begin{equation}
        B_{2} \rightarrow B_{2}-n\pi\text{Vol}(S^{2}_{1}).
    \end{equation}
For the NS5 charge, we integrate over the submanifold $\Sigma_{1}=(r,S^{2}_{1})$ for $\mu=0$ and $\mu=\pi$
    \begin{equation}
        N_{NS5}=\frac{1}{(2\pi)^{2}}\int_{\Sigma_{1}|_{\mu=0,\pi}}H_{3}=1.
    \end{equation}
This suggests that the location of the NS5-branes correspond to the location of the singularities of the background.\\
For the D5 charge, we integrate over the submanifold $\Sigma_{2}=(r,S^{2}_{2})$ for any value of $\mu$
    \begin{equation}
        N_{D5}=\frac{1}{(2\pi)^{2}}\int_{\Sigma_{2}}F_{3}=(2n+1)M.
    \end{equation}
Finally, for the D3 charge, we integrate over the cycle $\Sigma_{3}=(\mu,S^{2}_{1},S^{2}_{2})$ for any value of $r$
    \begin{equation}
        N_{D3}=-\frac{1}{(2\pi)^{4}}\int_{\Sigma_{3}}\widehat{F}_{5}=Mn.
    \end{equation}
    
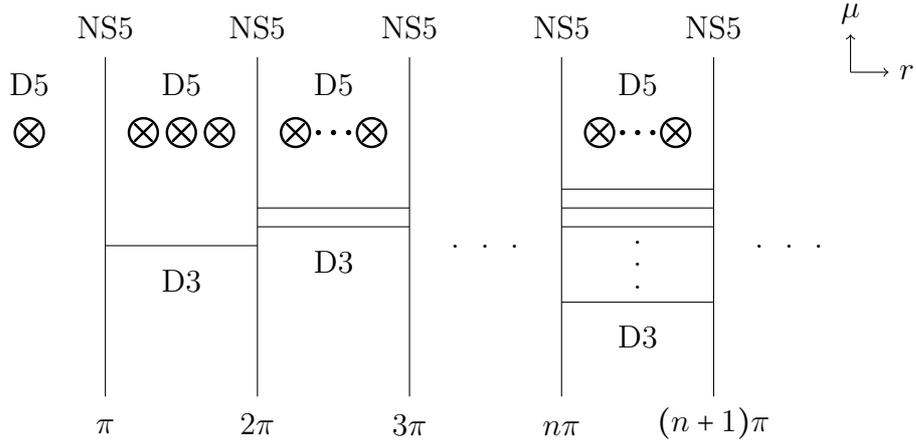
\begin{figure}[h!]
\begin{center}
\begin{tikzpicture}

\draw[->] (10.8,2.3) -- (10.8,2.8) node[above] {$\mu$};
\draw[->] (10.8,2.3) -- (11.3,2.3) node[right] {$r$};


\node[label=above:D5] at (0,1.5){\LARGE $\otimes$};
\node[label=above:] at (1.5,1.5){\LARGE $\otimes$};
\node[label=above:D5] at (2,1.5){\LARGE $\otimes$};
\node[label=above:] at (2.5,1.5){\LARGE $\otimes$};
\node[label=above:] at (3.5,1.5){\LARGE $\otimes$};
\node[label=above:] at (4,1.5){\LARGE $...$};
\node[label=above:] at (4.5,1.5){\LARGE $\otimes$};
\draw (4,1.85) node[above] {D5};
\node[label=above:] at (7.5,1.5){\LARGE $\otimes$};
\node[label=above:] at (8,1.5){\LARGE $...$};
\node[label=above:] at (8.5,1.5){\LARGE $\otimes$};
\draw (8,1.85) node[above] {D5};


\draw (1,-2) -- (1,2.5);
\draw (1,2.6) node[above] {NS5};
\draw (1,-2.16) node[below] {$\pi$};
\draw (3,-2) -- (3,2.5);
\draw (3,2.6) node[above] {NS5};
\draw (3,-2.09) node[below] {$2\pi$};
\draw (5,-2) -- (5,2.5);
\draw (5,2.6) node[above] {NS5};
\draw (5,-2.1) node[below] {$3\pi$};
\draw (7,-2) -- (7,2.5);
\draw (7,2.6) node[above] {NS5};
\draw (7,-2.19) node[below] {$n\pi$};
\draw (9,-2) -- (9,2.5);
\draw (9,2.6) node[above] {NS5};
\draw (9,-2) node[below] {$(n+1)\pi$};


\draw (1,0) -- (3,0);
\draw (3,0.5) -- (5,0.5);
\draw (3,0.25) -- (5,0.25);
\draw (5.4,0) node[right] {.~~.~~.};
\draw (7,0.75) -- (9,0.75);
\draw (7,0.5) -- (9,0.5);
\draw (7,0.25) -- (9,0.25);
\node at (8,0.05){$.$};
\node at (8,-0.25){$.$};
\node at (8,-0.55){$.$};
\draw (7,-0.75) -- (9,-0.75);
\draw (9.4,0) node[right] {.~~.~~.};
\draw (2,-0.75) node[above] {D3};
\draw (4,-.5) node[above] {D3};
\draw (8,-1.5) node[above] {D3};
\end{tikzpicture}
\end{center}
\caption{The Hanany-Witten (NS5, D3, D5) brane set-up for the NATD solution (taking $M=1$), in which $Mn$ D3-branes are stretched between two NS5 branes located at $n\pi$ and $(n+1)\pi$, with $(2n+1)M$ D5-branes in each interval.}
\label{fig:BraneSetup}
\end{figure}

\begin{figure}[h!]
\begin{center}
 \begin{tikzpicture}[square/.style={regular polygon,regular polygon sides=4}]
        \node at (0,0) [square,inner sep=0.3em,draw] (f100) {$M$};
        \node at (2,2) [square,inner sep=0.1em,draw] (f200) {$3M$};
        \node at (4,2) [square,inner sep=0.1em,draw] (f300) {$5M$};
        \node at (6,2) [square,inner sep=0.1em,draw] (f400) {$7M$};
        \node at (2,0) [circle,inner sep=0.5em,draw] (c100) {$M$};
        \node at (4,0) [circle,inner sep=0.4em,draw] (c200) {$2M$};
        \node at (6,0) [circle,inner sep=0.4em,draw] (c300) {$3M$};
        \draw (c100) -- (f200);
        \draw (c100) -- (c200);
        \draw (c200) -- (f300);
        \draw (c200) -- (c300);
        \draw (c300) -- (f400);
        \draw (c300) -- (7.5,0);
        \draw (8,0) node[right] {.~~.~~.};
    \end{tikzpicture}
    \end{center}
\caption{The corresponding overbalanced linear quiver diagram for the NATD geometry.}
\label{fig:Quiver2}
\end{figure}
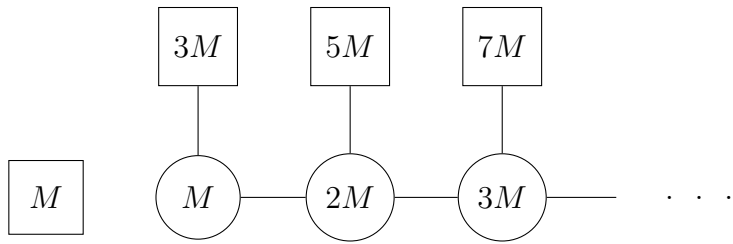

The full brane configuration of the entire geometry (for the case $M=1$) is described in figure
\ref{fig:BraneSetup}, in which $Mn$ D3-branes are stretched between two NS5-branes located at $n\pi$ and $(n+1)\pi$, with $(2n+1)M$ D5-branes in each interval. The corresponding (overbalanced) linear quiver diagram is then given in figure \ref{fig:Quiver2}.

\newpage

\section{On the Completion of Geometries}

\subsection{NATD as a zoom-in}

In \cite{Legramandi:2021uds}, by using the electrostatic problem formalism, evidence was presented on how the non-Abelian T-dual of AdS$_{6}$ in massive Type IIA backgrounds could be thought of as a particular zoom-in of a more general solution, which possesses a well-defined holographic dual. Here, we give a similar argument for the NATD of AdS$_{4}$ in Type IIA Supergravity.

In \cite{Akhond:2021ffz}, the infinite family of backgrounds dual to balanced linear quivers is derived from the following potential (we consider $\sigma>0$ for simplicity)
    \begin{equation}\label{LinearQuivers}
        V_{\text{LQ}}(\sigma,\eta) = \sum^{+\infty}_{k=0} \frac{a_{k}}{\sigma} \cos\left( \frac{k\pi\eta}{P} \right) e^{-\frac{k\pi\sigma}{P}},
    \end{equation}

with $a_{k}$ the coefficients of the Fourier expansion of the respective Rank function (which contains the ranks of colour and flavour groups). Expanding this potential in a Taylor series gives
    \begin{equation}
        V_{\text{LQ}}(\sigma,\eta) 
        = \frac{1}{\sigma} \sum^{+\infty}_{k=0}a_{k} 
        - \frac{\pi}{P} \sum^{+\infty}_{k=0}k\, a_{k} 
        + \frac{\pi^{2}}{2P^{2}}\left( -\frac{\eta^{2}}{\sigma} + \sigma \right) \sum^{+\infty}_{k=0}k^{2}\, a_{k}
        + \frac{\pi^{3}}{6P^{3}}\left( 3\eta^{2}-\sigma^{2} \right)\sum^{+\infty}_{k=0}k^{3}\, a_{k}
        + ...
    \end{equation}

By defining
    \begin{equation}
        \beta_{1} = \sum^{+\infty}_{k=0}a_{k}, \, \, \,\,
        \beta_{2} = - \frac{\pi}{P}\sum^{+\infty}_{k=0}k\, a_{k},\,\,\,\,
        \beta_{3} = -\frac{1}{3}\frac{\pi^{2}}{2P^{2}}\sum^{+\infty}_{k=0}k^{2}\, a_{k}, \,\,\,\,
        \beta_{4} = -\frac{1}{4}\frac{\pi^{3}}{6P^{3}}\sum^{+\infty}_{k=0}k^{3}\, a_{k},
    \end{equation}
one can check that the truncated expansion completely contains the potential for the NATD ($\beta_{1}$ does not appear on any background fields nor in the Rank function, and $\beta_{2}$ appears as a gauge transformation of $C_{2}$). As for the $\text{AdS}_6$ case, this suggest that the $\text{AdS}_4$ NATD corresponds to a zoom-in of a background for which we can completely characterise the holographic dual. This means that even though the NATD holographic dual is not properly defined, it can be completed into a 3d $\mathcal{N}=4$ linear quiver.

\subsection{Comments on the ATD}\label{ATD general solution}

The ATD background can be found via a similar zoom-in in the family of orthogonal functions to \eqref{LinearQuivers}, namely
    \begin{equation}\label{eqn:ATDzoomin}
        V_{\perp}(\sigma,\eta) = \sum^{+\infty}_{k=0} \frac{a_{k}}{\sigma} \sin\left( \frac{k\pi\eta}{P} \right) e^{-\frac{k\pi\sigma}{P}}.
    \end{equation}

Notice that, although this is a solution to the equation of motion, it does not satisfy the boundary conditions of the electrostatic problem. This means that it is not possible to use the same arguments to suggest that the ATD be completed into a more general geometry that has a well defined holographic dual. This is consistent with the fact that the ATD already has a precise holographic dual: the same one as its seed background. 

\section{A New Supergravity Solution}

In this section we investigate the background obtained when using a third finite part of the infinite Taylor series expansion of $V(\sigma,\eta)$. Note that, in addition to the previous two examples, it is the only other solution for which the $\text{AdS}_4$ warp factor $f_1(\sigma,\eta)$ depends only on a single coordinate. For a constant $f_1(\sigma,\eta)$, the string background was shown to be integrable for the case of $\text{AdS}_7$ in \cite{Filippas:2019puw} (where the classical 2d dynamics of the $\sigma$-model can be written in a Lax pair leading to an infinite number of conserved charges). However, in this case, a constant $f_1(\sigma,\eta)$ is not possible. Hence, an $\text{AdS}_4$ warp factor of a single coordinate is the next best option, as integrability may occur for fixed values of the coordinate. 

Now we present the new solution of Type IIB Supergravity that comes from considering the following higher order truncation to equation \eqref{eqn:ATDzoomin}
    \begin{equation}\label{eqn:new}
        V_\text{HO} = \alpha_{4}\left( 4\frac{\eta^{3}}{\sigma}-12\eta\sigma \right).
    \end{equation}
Consider the change of coordinates
    \begin{align}
        \sigma &= \frac{2\sqrt{2}}{\pi}~r\sin(\theta),\\
        \eta &= \frac{2\sqrt{2}}{\pi}~r\cos(\theta).
    \end{align}
After fixing $\alpha_{4} = M/64$, we find
    \begin{equation}
        ds^{2} =  r\left( ds^{2}(\text{AdS}_{4}) 
        + \frac{\cos(2\theta)}{\tilde{\Delta}}ds^{2}(S^{2}_{1})
        + 2\sin^{2}\left(\theta\right) ds^{2}(S^{2}_{2}) 
        + \frac{2}{\cos(2\theta)}\left(\frac{dr^{2}}{r^2}+d\theta^{2}\right)\right), 
    \end{equation}
where $\tilde{\Delta} = 2\cos^{2}(\theta)+1$. In order to ensure that the metric does not change signature, we must make the restriction $\theta \in  \left[0, \frac{\pi}{4} \right]$. 

In addition, there is the Dilaton
    \begin{equation}
        e^{-2\Phi} = \frac{9M^2}{\pi^2} r^{2}\cos(2\theta)\tilde{\Delta},
    \end{equation}
and the NS and R forms
    \begin{align}
        B_{2} &= 2\sqrt{2}\,\frac{r\cos(\theta)}{\tilde{\Delta}} \text{Vol}(S^{2}_{1}), \\
        C_{2} &=  \frac{12}{\pi}M r^{2}\sin^{3}(\theta)\cos(\theta) \text{Vol}(S^{2}_{2}),\\
        \tilde{C}_{4} &= \frac{6\sqrt{2}}{\pi}Mr^{3}\cos(\theta)\text{Vol}(\text{AdS}_{4}),
    \end{align}
which lead to the following field strengths
    \begin{align}
    H_{3} &= \frac{\sqrt{2}}{\tilde{\Delta}^{2}}\bigg( 2 \cos(\theta)\tilde{\Delta}\,dr 
             + \,r\,\Big(\sin(3\theta)-\sin(\theta)\Big)\, d\theta \bigg)\wedge\text{Vol}(S^{2}_{1})  ,\\
    F_{3} &= \frac{M}{\pi}\Big( 24 \sin^{3}(\theta)\cos(\theta)\, r dr 
             + 12 r^{2}(2\tilde{\Delta}-3)\sin^{2}(\theta)\, d\theta\Big)\wedge \text{Vol}(S^{2}_{2}),
    \end{align}
and also
    \begin{equation}
    \begin{aligned}
    F_{5} &= d\left( \frac{6\sqrt{2}}{\pi}Mr^{3}\cos(\theta) \right)\wedge\text{Vol}(\text{AdS}_{4})\\
          &\phantom{=} - \frac{12\sqrt{2}}{\pi \tilde{\Delta}}M\,r^{2}\cos(2\theta)\sin^{2}(\theta)
               \Big( \sin(\theta)\, dr + 3r\cos(\theta) \, d\theta  \Big)
               \wedge\text{Vol}(S^{2}_{1})\wedge \text{Vol}(S^{2}_{2}).
    \end{aligned}
    \end{equation}

This background has the nice property of a dilatation, in which, under the transformation $r \rightarrow \lambda r$ (where $\lambda$ is some constant), everything simply scales independently by differing powers of $\lambda$, leaving the forms of the background otherwise invariant. This is a slightly weaker condition to the ATD case, in which the background metric and field strengths are totally translation invariant. It is interesting to observe however, that the solution presented here, appearing as a higher order truncation of equation \eqref{eqn:ATDzoomin}, comes from the same family of solutions as the ATD.

\subsubsection*{Page Charges}

In the previous two solutions, the interval of $\mu$ was fixed by singularities, but we were free to choose the form of the interval of $r$ in order to ensure quantisation of the Page charges. In the NATD case, this included splitting $r \in [0,+\infty[$ into an infinite number of finite `minimal building block' intervals. The same needs to be done here, noting the presence of a singularity at $r=0$. Hence, choosing $\theta \in [0,\frac{\pi}{4}]$ and  $r \in [n\pi,(n+1)\pi]$, with $n \in \mathbb{N}$, integer quantisation of the Page charges can be achieved. \\\\
For the NS5 charge, we integrate in the submanifold $\chi_{1}=(r,S^{2}_{1})$, fixing $\theta=\pi/4$ to obtain integer quantisation (as taking $\theta=0$ leads to an irrational number of $N_{NS5}$ branes)
    \begin{equation}
         N_{NS5}=\frac{1}{(2\pi)^{2}}\int_{\chi_{1}|_{\theta=\pi/4}}H_{3}=1.
    \end{equation}
For the D5 charge, we first integrate in the submanifold $\chi_{2}=(r,S^{2}_{2})$, fixing $\theta=\pi/4$ (as $\theta=0$ gives zero charge)
 \begin{equation}
        N_{D5}=\frac{1}{(2\pi)^{2}}\int_{\chi_{2}|_{\theta=\pi/4}}F_{3}=3M(2n+1).
    \end{equation}
Finally, for the D3 charge, we integrate in the submanifold $\chi_{3}=(\theta,S^{2}_{1},S^{2}_{2})$ at fixed $r=n\pi$
    \begin{equation}
         N_{D3}=-\frac{1}{(2\pi)^{4}}\int_{\chi_{3}|_{r=n\pi}}\widehat{F}_{5}=2Mn^3.
    \end{equation}
     
\begin{figure}[h!]
\begin{center}
 \begin{tikzpicture}[square/.style={regular polygon,regular polygon sides=4}]
        \node at (0,0) [square,inner sep=0.1em,draw] (f100) {$3M$};
        \node at (2,2) [square,inner sep=0.1em,draw] (f200) {$9M$};
        \node at (4,2) [square,inner sep=-0.15em,draw] (f300) {$15M$};
        \node at (6,2) [square,inner sep=-0.15em,draw] (f400) {$21M$};
        
        \node at (2,0) [circle,inner sep=0.4em,draw] (c100) {$2M$};
        \node at (4,0) [circle,inner sep=0.25em,draw] (c200) {$16M$};
        \node at (6,0) [circle,inner sep=0.25em,draw] (c300) {$54M$};
        \draw (c100) -- (f200);
        \draw (c100) -- (c200);
        \draw (c200) -- (f300);
        \draw (c200) -- (c300);
        \draw (c300) -- (f400);
        \draw (c300) -- (7.5,0);
        \draw (8,0) node[right] {.~~.~~.};
    \end{tikzpicture}
    \end{center}
\caption{The corresponding overbalanced linear quiver diagram for the new geometry.}
\label{fig:Quiver3}
\end{figure}
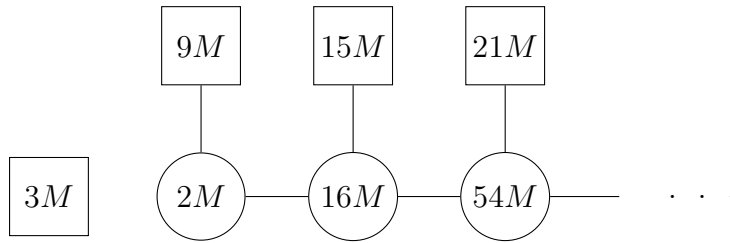

 The brane configuration for this geometry is shown in figure \ref{fig:3BraneSetup}, in which $2Mn^3$ D3-branes are stretched between two NS5-branes located at $n\pi$ and $(n+1)\pi$, with $3M(2n+1)$ D5-branes in each interval. The corresponding (overbalanced) linear quiver diagram is then given in figure \ref{fig:Quiver3}.

\begin{figure}[h!]
\begin{center}
\begin{tikzpicture}

\draw[->] (11.5,2.3) -- (11.5,2.8) node[above] {$\theta$};
\draw[->] (11.5,2.3) -- (12,2.3) node[right] {$r$};


\node[label=above:] at (0,1.5){\LARGE $\otimes$};
\node[label=above:D5] at (0.5,1.5){\LARGE $\otimes$};
\node[label=above:] at (1,1.5){\LARGE $\otimes$};
%
\node[label=above:] at (2,1.5){\LARGE $\otimes$};
\node[label=above:] at (2.5,1.5){\LARGE $...$};
\node[label=above:] at (3,1.5){\LARGE $\otimes$};
\draw (2.5,1.85) node[above] {D5};
\node[label=above:] at (4,1.5){\LARGE $\otimes$};
\node[label=above:] at (4.5,1.5){\LARGE $...$};
\node[label=above:] at (5,1.5){\LARGE $\otimes$};
\draw (4.5,1.85) node[above] {D5};
\node[label=above:] at (8,1.5){\LARGE $\otimes$};
\node[label=above:] at (8.5,1.5){\LARGE $...$};
\node[label=above:] at (9,1.5){\LARGE $\otimes$};
\draw (8.5,1.85) node[above] {D5};


\draw (1.5,-2) -- (1.5,2.5);
\draw (1.5,2.6) node[above] {NS5};
\draw (1.5,-2.16) node[below] {$\pi$};
\draw (3.5,-2) -- (3.5,2.5);
\draw (3.5,2.6) node[above] {NS5};
\draw (3.5,-2.09) node[below] {$2\pi$};
\draw (5.5,-2) -- (5.5,2.5);
\draw (5.5,2.6) node[above] {NS5};
\draw (5.5,-2.1) node[below] {$3\pi$};
\draw (7.5,-2) -- (7.5,2.5);
\draw (7.5,2.6) node[above] {NS5};
\draw (7.5,-2.19) node[below] {$n\pi$};
\draw (9.5,-2) -- (9.5,2.5);
\draw (9.5,2.6) node[above] {NS5};
\draw (9.5,-2) node[below] {$(n+1)\pi$};


\draw (1.5,0) -- (3.5,0);
\draw (1.5,-0.25) -- (3.5,-0.25);
\draw (3.5,0.75) -- (5.5,0.75);
\draw (3.5,0.5) -- (5.5,0.5);
\draw (3.5,0.25) -- (5.5,0.25);
\node at (4.5,0.05){$.$};
\node at (4.5,-0.25){$.$};
\node at (4.5,-0.55){$.$};
\draw (3.5,-0.75) -- (5.5,-0.75);

\draw (5.9,0) node[right] {.~~.~~.};
\draw (7.5,0.75) -- (9.5,0.75);
\draw (7.5,0.5) -- (9.5,0.5);
\draw (7.5,0.25) -- (9.5,0.25);
\node at (8.5,0.05){$.$};
\node at (8.5,-0.25){$.$};
\node at (8.5,-0.55){$.$};
\draw (7.5,-0.75) -- (9.5,-0.75);
\draw (9.9,0) node[right] {.~~.~~.};

\draw (2.5,-1) node[above] {D3};
\draw (4.5,-1.5) node[above] {D3};
\draw (8.5,-1.5) node[above] {D3};

\end{tikzpicture}
\end{center}
\caption{The Hanany-Witten (NS5, D3, D5) brane set-up for the new solution, in which $2Mn^3$ D3-branes are stretched between two NS5-branes located at $n\pi$ and $(n+1)\pi$, with $3M(2n+1)$ D5-branes in each interval (taking $M=1$ in the figure).}
\label{fig:3BraneSetup}
\end{figure}
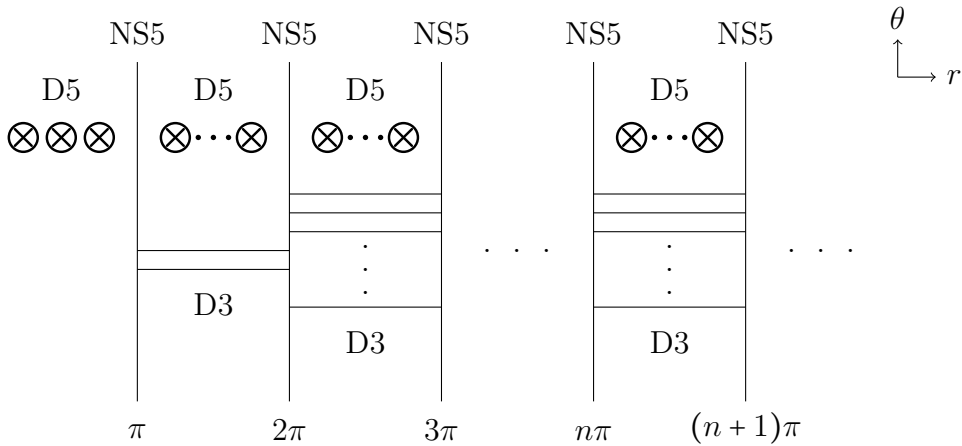

 \newpage
\section{Conclusions}
In this paper we used the electrostatic problem formalism developed in \cite{Akhond:2021ffz} to obtain three AdS$_{4}\times S^{2}\times S^{2}$ backgrounds in Type IIB Supergravity, derived from the polynomial solutions of the 2D Laplace equation defining the geometry. However, they do not satisfy the boundary conditions that lead to linear quiver field theories. Two of these solutions correspond to the Abelian and non-Abelian T-duals of the Type IIA background obtained from the dimensional reduction of AdS$_{4}\times S^{7}$. We therefore re-derive the NATD solution using a different approach to \cite{Lozano:2016wrs}, characterising the solutions by specific terms in the polynomial expansion of the potential. Hence, the dual Field Theory should be more easily determined via this approach. The third solution is a new Supergravity background, which in addition to the ATD and NATD cases, form the three backgrounds with the simplest possible $\text{AdS}_4$ warp factor - depending on only a single coordinate. We then computed the Page charges for each background, and described the Hanany-Witten brane set-up and linear quiver diagrams for all three geometries. As in previous references, we find that the NATD brane set-up is unbounded, making the dual field theory description obscure. 


We then showed how the potential leading to the NATD solution emerges as a truncation of the full solution of the electrostatic problem, which leads to linear quiver field theories as holographic duals. This agrees with the picture that NATD solutions can be completed into a background that has a well-defined holographic dual. The ATD backgrounds are self-consistent on their own, i.e. since they describe the same dynamics as their seed, they are dual to the same field theory. The new background has a nice dilatation and originates from the same general solution as the ATD case.

Interestingly, each term in the expansion of the potential is in itself a solution, meaning any unique combination will lead to a different background - with the ATD and NATD cases emerging naturally. The dynamics of the full $\text{AdS}_4$ Type IIB Supergravity solution dual to the linear quivers is then constructed from the separate constituent dynamics of each term in the Taylor series. 

It is worth noting that, although they are obtained as a truncation of different families of solutions, the NATD and the new solution have a brane set-up that are unbounded. It would be interesting to see whether all other solutions that can be obtained as a truncation share this same property. In the same line, if one were to find a solution coming from a higer order term in the Taylor series expansion of the potential that leads to the linear quiver, one could use the same arguments as for the NATD to argue that those solutions can be completed. This is not true for the solutions that belong to the same family as the ATD, such as the new background presented here. Since these solutions come from a potential that does not satisfy the boundary conditions of the electrostatic problem, it is not possible to argue that they can be completed in a similar way, with the ATD being an exception to this argument. It is then yet to be seen whether those solutions can be completed into a background with a well-defined holographic dual.

\section*{Acknowledgments}

The authors thank Mohammad Akhond, Andrea Legramandi, Carlos Nuñez and Lucas Schepers for useful discussions and guidance. The work of R.S. is supported by STFC grant ST/W507878/1.

\newpage
\appendix

\end{document}